\documentclass[prc,preprint,showpacs,showkeys,nofootinbib,tightenlines]{revtex4}

\usepackage{graphicx}  %
\usepackage{bm}  %                                          

\newcommand{\simle}{\hspace*{0.2em}\raisebox{0.5ex}{$<$}
     \hspace{-0.8em}\raisebox{-0.3em}{$\sim$}\hspace*{0.2em}}

\newcommand{\beq}{\begin{equation}}
\newcommand{\eeq}{\end{equation}}
\newcommand{\bqa}{\begin{eqnarray}}
\newcommand{\eqa}{\end{eqnarray}}

\def\mqo2{{\!\!\!}}

\begin{document}

\title{Universality in the Three-Body Problem for $^4$He Atoms}

\author{Eric Braaten}\email{braaten@mps.ohio-state.edu}
\author{H.-W. Hammer}\email{hammer@itkp.uni-bonn.de}\thanks{Present address: 
Helmholtz-Institut f\"ur Strahlen- und Kernphysik (Theorie), Universit\"at 
Bonn, Nussallee 14-16, 53115 Bonn, Germany}

\affiliation{Department of Physics,
         The Ohio State University, Columbus, OH\ 43210, USA}

%\date{\today}
\date{December 18, 2002}

\begin{abstract}
The two-body scattering length $a$ for $^4$He atoms is much larger than 
their effective range $r_s$. As a consequence, low-energy few-body
observables 
have universal characteristics that are independent of the interaction
potential. Universality implies that, up to corrections
suppressed by $r_s/a$, all low-energy three-body
observables are determined by $a$ and a three-body parameter 
$\Lambda_*$. We give simple expressions
in terms of $a$ and $\Lambda_*$ for the trimer binding energy equation,
the atom-dimer scattering phase shifts, and the rate for three-body 
recombination at threshold. We determine $\Lambda_*$ for several 
$^4$He potentials from the calculated binding energy of the
excited state of the trimer and use it to obtain the universality
predictions for the other low-energy observables. We also use the calculated
values for one potential to estimate the effective range corrections
for the other potentials.
\end{abstract}

\smallskip
\maketitle

\section{Introduction}
The interactions of nonrelativistic particles with extremely low energies, 
such as cold atoms, are determined primarily
by their S-wave scattering length $a$. 
Typically $|a|$ is of the order of the natural length scale $l$ associated with
low-energy interactions, which for short-range interactions is given
by the range of the potential. If $|a|$ is much larger than $l$, however, 
low-energy atoms exhibit universal properties that 
are independent of the interaction potential.
In the two-body sector, universality implies that low-energy observables
are determined by the single parameter $a$  up to corrections 
suppressed by $l/|a|$. In particular, the cross 
section for low-energy atom-atom scattering
is a simple function of $E$ and $a$ only.
If $a>0$, there is also a shallow two-body bound state
(the dimer) whose binding energy is determined by $a$:
$B_2=\hbar^2/ma^2$, where $m$ is the mass of the atoms \cite{textbook}.

Efimov showed that if the two-body scattering length is large,
there are also universal properties
in the three-body sector \cite{Efi71}. The most remarkable is a 
sequence of three-body bound states (trimers) with binding energies 
geometrically spaced in the interval 
between $\hbar^2/ma^2$ and $\hbar^2/ml^2$. 
In addition to the binding energies of these Efimov trimers, the
low-energy three-body observables include scattering rates 
for three atoms and, if $a>0$, an atom and a dimer. The consequences
of universality in the three-body sector are simplest if the 
potential supports no deep two-body bound states with
binding energies of order $\hbar^2/ml^2$. In this case, the low-energy
three-body observables are determined by $a$ and a single
three-body parameter up to corrections suppressed
by $l/|a|$. A simple physical observable that can be used as the 
three-body parameter is the binding energy of the Efimov state closest
to threshold. Alternatively, the 
three-body parameter can be specified by a
boundary condition on the three-body wave function at short
distances. In Ref.~\cite{BHK99}, the authors introduced a more abstract
three-body parameter $\Lambda_*$
with dimensions of wave number (defined in Eq.~(\ref{B3-Efimov}) below)
which is particularly convenient for quantitative
calculations within the effective field theory approach.

A large two-body scattering length can be obtained by fine-tuning a
parameter in the interatomic potential to bring a 
real or virtual two-body bound state close to the 2-atom threshold. 
The fine tuning can be obtained experimentally by adjusting an
external variable, such as a magnetic field \cite{Festh} or an electric 
field \cite{NFJ99}. Large scattering lengths for $^{23}$Na and
$^{85}$Rb atoms have been obtained in the laboratory by 
tuning an external magnetic field to the neighbourhood 
of a Feshbach resonance \cite{Fesex}.
The fine tuning can also be provided accidentally by nature. 
A prime example is the $^4$He atom, whose scattering length $a\approx
104$ \AA\ \cite{Gri00} is much larger than its effective range $r_s 
\approx 7.3$ \AA\ \cite{JaA95} which can be taken as an estimate of the
natural low-energy length scale $l$.

The large scattering length for $^4$He makes this atom an ideal
example of universality. The experimental information on 
low-energy $^4$He atoms is rather limited.
Using diffraction of a molecular beam of small $^4$He clusters from a 
transmission grating, the bond length of the $^4$He dimer has recently 
been measured  \cite{Gri00}: $\langle r\rangle= (52 \pm 4)$ \AA,
which is an order magnitude larger than the effective range.
The scattering length $a=\left(104^{+8}_{-18}\right)$\AA\ and the
dimer binding energy $B_2=\left(1.1^{+0.3}_{-0.2}\right)$ mK
were derived from the measured bond length using the zero range 
approximation \cite{Gri00}. The $^4$He trimer and several larger 
$^4$He clusters have been observed \cite{STo96,BST02}, but
no quantitative experimental information about their binding energies
is available to date. However, there have been extensive theoretical
calculations of the few-body problem for $^4$He using modern
two-body potentials \cite{LM2M2,TTY}. Theoretical calculations of trimer
binding energies have also improved, so that they now have several 
digits of accuracy \cite{NFJ98,RoY00,MSSK01,Lew97}.
They indicate that there are 
two trimers: a ground state with binding energy $B_3^{(0)}$ and an 
excited state with binding energy $B_3^{(1)}$. The ground state
binding energies of larger $^4$He clusters have been calculated
using the diffusion Monte Carlo method \cite{Lew97}.
Their excited state binding energies
have been calculated using a combination of Monte Carlo methods and 
the hyperspherical adiabatic approximation \cite{BlG00}.
There have also been some calculations of three-body scattering 
observables. The S-wave phase shifts for atom-dimer scattering 
have been calculated in Refs.~\cite{BHK99,MSSK01}.
The three-body recombination rate has been calculated at threshold 
\cite{FRS96,NiM99,EGB99,BBH00} and as a function of the energy \cite{SEG02}.

Universality implies that in the limit of large scattering
length all low-energy three-body observables are determined by the 
scattering length $a$ (or equivalently the dimer binding energy
$B_2$) and one three-body parameter such as $\Lambda_*$. In order
to determine this parameter, a low-energy three-body observable
is required as input. We can take advantage of the accurate 
calculations of the trimer binding energies for modern $^4$He 
potentials by using $B_3^{(1)}$
to determine $\Lambda_*$ for $^4$He. Once this parameter
is determined, universality can be used to predict all other low-energy
three-body observables for $^4$He atoms.

We can also exploit accurate theoretical calculations for older
$^4$He potentials to demonstrate the nontrivial nature of
universality in the three-body sector \cite{FTD99,DFT00}. 
Scaling variables are dimensionless combinations of physical observables.
Universality implies that three-body scaling
variables are nontrivial universal functions of $a\Lambda_*$.
By eliminating the dependence on $\Lambda_*$, we can express 
one scaling variable as a universal function of any other scaling
variable. The various $^4$He potentials have slightly different scattering
lengths $a$ and also different values of $\Lambda_*$.
Therefore, if one scaling variable is plotted as a function of a second,
the points for various $^4$He potentials should all lie along a universal
line. Frederico, Tomio, Delfino, and Amorim calculated
the scaling function relating $B_3^{(1)}/B_3^{(0)}$ 
to $B_2/B_3^{(0)}$ and showed that the points for various $^4$He
potentials lie close to the universal scaling curve \cite{FTD99,DFT00}.

In this paper, we collect all the information that is currently available 
on the universal properties in the three-body system of $^4$He atoms. 
We go beyond the work in Refs.~\cite{FTD99,DFT00} in various respects.
We give explicit parametrizations in terms of $a$ 
and the three-body parameter $\Lambda_*$ for many low-energy 
three-body observables, including Efimov binding energies,
atom-dimer scattering phase shifts, and the three-body recombination rate.
We also calculate additional scaling functions
and estimate the effective range corrections.
In Section \ref{sec:2bdy}, we briefly review universality in the 
two-body sector. In Section \ref{sec:bound},
we discuss the universal properties of the trimer binding energies.
We determine the three-body parameter $\Lambda_*$
for various $^4$He potentials from the excited state
binding energy $B_3^{(1)}$ and use universality to
predict the ground state energy $B_3^{(0)}$. We also 
demonstrate universality by exhibiting the correlation between the 
scaling variables $B_3^{(0)}/B_2$ and $B_3^{(1)}/B_2$ for
various $^4$He potentials.
In Sections \ref{sec:2plus1} and \ref{sec:3brec},
we discuss the universal properties in atom-dimer scattering and
three-body recombination, respectively. We use universality to 
predict the S-wave scattering length $a_{12}$ and effective
range $r_{s,12}$ for atom-dimer scattering as well as the rate 
constant for three-body recombination at threshold. We also 
demonstrate universality by exhibiting
the correlations between the scaling variable $a_{12}/a$ and the
energy scaling variables for various $^4$He potentials.
Section \ref{sec:conc} contains a summary and the outlook for using
universality as the basis for a quantitative description of cold atoms.

\section{Two-Body Sector}
\label{sec:2bdy}

We begin by reviewing the universal properties in the two-body
sector for $^4$He \cite{textbook}. 
We use the phrase \lq\lq low-energy'' to
refer to energies close to the threshold for free atoms.
There is a natural length scale $l$ for low-energy
observables. For a short-range potential, $l$ is set by the
range. If the potential has a van der Waals tail, $V(r)\to -C_6/r^6$,
the natural low-energy length scale is $l\approx (mC_6/\hbar^2)^{1/4}$.
The natural energy scale for low-energy observables is $\hbar^2/ml^2$.
For $^4$He, $C_6=1.46$ a.u. which leads to  
$l=5$ \AA\ and $\hbar^2/ml^2=0.4$ K. 
Universality occurs because a parameter in the 
two-body potential has been tuned such that the scattering length
is unnaturally large. The scattering length $a= 104$ \AA\ \cite{Gri00} 
for $^4$He is much larger than $l$. 
We can interpret the
large scattering length as the result of an accidental fine tuning
by nature of either a parameter in the two-body potential, such as its
overall strength, or of the mass of the $^4$He atom. 
The $^3$He atom has a mass that is only about $3/4$ of that of $^4$He, 
and its scattering length is $-7.1$ \AA\ \cite{LM2M2},
close in magnitude to the natural low-energy length scale. 

If $|a| \gg l$, universality can be used to describe the low-energy 
observables for atoms in spite of the fact that the van der Waals tail 
makes the potential long-range. 
Explicit expressions for the scattering length and effective range
for a potential with a van der Waals tail have been derived in 
Ref.~\cite{Flam93}. 
A long-range potential introduces nonanalytic behavior in
the dependence of the scattering amplitudes on the wavevector ${\bf k}$.
However if the potential falls off like $1/r^6$, this nonanalytic behavior
enters first at 4$^{\rm th}$ order in $k$.  Such effects can not be 
reproduced by a short-range potential.  Fortunately, their effects on 
low-energy observables are suppressed by 4 powers of $l/a$.
We will focus on the universality predictions at leading order in $l/a$
and also on the effective range corrections that are first order in $l/a$.
At this level of accuracy, the effects of the van der Waals tail 
on low-energy observables can be reproduced by a short-range potential.
Realistic interatomic potentials will therefore exhibit the same
universal characteristics as short-range potentials. 

The two-body observables are the differential cross sections for 
two-body scattering and the binding energies for two-body
bound states. The
differential cross section for the elastic scattering of two
identical spinless bosons with total energy $E=\hbar^2 k^2/m$
has the general form
\beq
\frac{d\sigma}{d\Omega}=4\left|\sum_{L\ {\rm even}}
\frac{2L+1}{k\cot\delta_L(k)-ik} P_L(\cos\theta)\right|^2\,,
\label{eq:dsigatoms}
\eeq
where $\delta_L(k)$ is the phase shift for the $L$'th partial
wave. The total cross section is obtained from Eq.~(\ref{eq:dsigatoms})
by integrating over a solid angle of $2\pi$ to avoid counting
identical particles twice.
At low energies, the cross section is dominated by the 
$L=0$ \lq\lq S-wave'' term. The effective range expansion of 
$k\cot\delta_0(k)$ has the form
\beq
\label{eq:ere}
k\cot\delta_0(k)=-\frac{1}{a}+\frac{1}{2}r_s k^2+\ldots\,.
\eeq
The first two coefficients define the scattering length $a$ and the
effective range $r_s$. The natural size for these coefficients is the 
low-energy length scale $l$. The binding energies of the two-body
bound states are determined by the poles of the scattering amplitude.
For S-wave bound states, the binding energies
are $B_2^{(n)}=\hbar^2\kappa_n^2/m$, where $\kappa_n$ is a
solution to the equation
\beq
\label{eq:2bdybound}
i\kappa\cot\delta_0(i\kappa)+\kappa=0\,.
\eeq
The natural energy scale for a bound state close to threshold
is $\hbar^2/ml^2$.

If the scattering length $a$ is unnaturally large, the differential 
cross section exhibits universal behavior at energies 
small compared to $\hbar^2/ml^2$:
\beq
\label{dsig:uni}
\frac{d\sigma}{d\Omega}=\frac{4a^2}{1+k^2 a^2}\,,\qquad 
kl \ll 1,\quad |a| \gg l\,.
\eeq
The leading correction comes from the effective range.
At wave number $k$ of order $(r_s a)^{-1/2}$ or smaller, the error is of
order $r_s/a$. At larger wave numbers, the error increases like
$k^2 r_s^2$ and becomes of order one at $k$ of order $1/r_s$. 
Note that the differential cross section (\ref{dsig:uni})
is determined as a function of $k$ by the single parameter $a$.

If $a$ is large and positive, there is one additional low-energy observable.
There is a shallow two-body S-wave bound state that we will
refer to as \lq\lq the dimer''. Up to corrections suppressed
by $l/a$, its binding energy is
\beq
B_2=\frac{\hbar^2}{ma^2}\,,\qquad   a\gg l\,.
\label{eq:B2}
\eeq
Alternatively, if we take $B_2$ as input, universality 
gives a prediction for the scattering length:
\beq
a_B \equiv\frac{\hbar}{\sqrt{mB_2}}\,.
\label{eq:ab}
\eeq
The leading correction to the universal prediction for 
$B_2$ in (\ref{eq:B2}) comes from the effective range.
If we truncate the effective range expansion (\ref{eq:ere})
after the $k^2$ term, Eq.~(\ref{eq:2bdybound}) is
a quadratic equation with two solutions:
\beq
\label{eq:B2range}
B_2^{(\pm )}=\frac{\hbar^2}{m}\frac{2}{r_s^2}\left[
1-\frac{r_s}{a}\pm \sqrt{1-2\frac{r_s}{a}}\,\right]\,.
\eeq
The solution $B_2^{(+)}$ is an artifact of the truncation.
We would expect a state with such an energy only if the
higher order terms in the effective range expansion are unnaturally
small. The solution $B_2^{(-)}$ is the binding energy of the shallow 
dimer. If we expand to first order in $r_s/a$, we obtain
\beq
B_2^{(-)}=\frac{\hbar^2}{ma^2}\left[1+\frac{r_s}{a}\right]\,.
\label{eq:B2rangeexp}
\eeq
The corrections from higher orders in the effective range
expansion (\ref{eq:ere}) are suppressed by $l^2/a^2$ and should 
therefore be comparable in magnitude to the $r_s^2/a^2$ correction.

We now consider the two-body observables for $^4$He. In Table~\ref{tab0},
we give the calculated scattering length \cite{MSSK01}, 
the effective range \cite{JaA95}, and the dimer binding energies 
\cite{MSSK01} for four commonly used potentials:
two modern potentials LM2M2 \cite{LM2M2} and TTY \cite{TTY}, and
two older potentials HFDHE2 \cite{HFDHE2} and HFD-B \cite{HFD-B}.
%%%%%%%%%%%%%%%%%%%%%%%%%%%%%%%%%%%%%%%%%%%%%%%%%%%%%%%%%%%%%%%%%%%%%%%%
\begin{table}[htb]
\begin{tabular}{c||c|c|c||c|c|c}
Potential & $a$ & $r_s$ & $B_2$ & $a_B$ & 
$\frac{\hbar^2}{ma^2}$ & $B_2^{(-)}$\\
\hline\hline
HFDHE2 & 124.65 & 7.396 & 0.83012 & 120.83 & 0.7800 & 0.8263 \\
HFD-B  & 88.50  & 7.277 & 1.68541 & 84.80  & 1.5474 & 1.6746 \\
LM2M2  & 100.23 & 7.326 & 1.30348 & 96.43  & 1.2064 & 1.2946 \\
TTY    & 100.01 & 7.329 & 1.30962 & 96.20  & 1.2117 & 1.3005
\end{tabular}
\caption{Two-body observables for four $^4$He potentials.
Lengths and energies are given in \AA\ and mK, respectively. The first
three columns show the calculated scattering lengths $a$ 
\protect\cite{MSSK01}, effective ranges $r_s$ \protect\cite{JaA95}, and 
dimer binding energies $B_2$ \protect\cite{MSSK01}. 
The last three columns show the universality predictions $a_B$ for $a$
using $B_2$ as input, the universality predictions for $B_2$ using $a$ 
as input, and the predictions for $B_2$ including the first-order 
effective range correction. 
(For $^4$He, the conversion constant is $\hbar^2/m =12.1194$ K$\,$\AA$^2$.)
}
\label{tab0}
\end{table}
%%%%%%%%%%%%%%%%%%%%%%%%%%%%%%%%%%%%%%%%%%%%%%%%%%%%%%%%%%%%%%%%%%%%%%%%  

In Table~\ref{tab0}, we also give some simple theoretical predictions
for two-body observables. We give the prediction $a_B$ for the 
scattering length obtained from (\ref{eq:ab}) by using $B_2$ as 
input. We also give two predictions
for the dimer binding energy $B_2$ using the scattering data $a$ and
$r_s$ as input. They are the universality prediction in 
Eq.~(\ref{eq:B2}) and the prediction including 
the first-order effective range correction
in Eq.~(\ref{eq:B2rangeexp}). We can obtain estimates of the theoretical
errors in approaches based on the universality at large $a$ by
comparing those approximations with the calculated value of $B_2$.
The universality prediction $\hbar^2/ma^2$ differs from $B_2$ by at
most 8\% and the errors decrease to at most 0.7\% if the
first order effective range correction is included.
This suggests that predictions of low-energy observables
based on universality should have an accuracy of about 10\% and that
that one may be able to reduce the errors to about 1\% by 
including effective range corrections.

\section{Trimer Binding Energies}
\label{sec:bound}

The most dramatic prediction of universality in the three-body sector
with large scattering length is the existence of Efimov states \cite{Efi71}.
They are a sequence of shallow three-body bound states with binding
energies much smaller than $\hbar^2/ml^2$. If a parameter in the
two-body potential is tuned such that $a\to \pm\infty$, the number of
these states increases roughly as $\ln (|a|/l)/\pi$. The spacing
of the deeper states is roughly geometric, with the ratio of
successive binding energies approaching 515.
The suggestion that the excited state of the $^4$He trimer is an
Efimov state was first made in Ref.~\cite{LDD77}.
Accurate calculations using modern potentials support that
interpretation \cite{NFJ98,MSSK01,RoY00,ELG96}.
In Refs.~\cite{BHK99,BHK02,FTD99}, it was argued that the trimer 
ground state is also an Efimov state.
If it is an Efimov state, universality can be used to predict its
binding energy. We will show that the resulting predictions are within
the errors expected from effective range corrections. We will also
give a definitive criterion for a three-body bound state to be an Efimov 
state.

Efimov derived some powerful constraints on low-energy three-body
observables for systems with large scattering length \cite{Efi71}.
They follow from the approximate scale-invariance at 
length scales $R$ in the region $l \ll R \ll |a|$ together with the 
conservation of probability. 
He introduced polar variables $H$ and $\xi$ in
the plane whose axes are $1/a$ and the energy variable
${\rm sgn}(E)|mE|^{1/2}/\hbar$. The angular variable $\xi$ is
\bqa
\xi  &=& -\arctan (a\sqrt {mB_3}/ \hbar)\,, \qquad\qquad\; a > 0 
\,,\nonumber\\
&=& -\pi + \arctan (|a|\sqrt {mB_3}/ \hbar)\,, \qquad a < 0 \,.
\label{Hxi-def}
\eqa
Efimov showed that low-energy three-body observables 
are determined by a few universal functions of the angle $\xi$.
In particular, the binding energies of the Efimov states are solutions to 
an equation involving a single universal function
$\Delta(\xi)$ \cite{Efi71}. 
Efimov's equation for the binding energies reads 
\cite{Efi71,BHK02}
\beq
B_3  + {\hbar^2 \over ma^2} = \frac{\hbar^2 \Lambda_*^2}{m}\,
e^{2 \pi n/ s_0} \exp \left[\Delta \left( \xi \right)/s_0
\right]\,,
\label{B3-Efimov}
\eeq
where $s_0\approx 1.00624$ is a transcendental number that
satisfies the equation
\beq
\sqrt{3}\,s_0 \cosh(\pi s_0/2)=8\sinh(\pi s_0/6)\,.
\eeq
We use a three-body parameter $\Lambda_*$ that was first introduced 
in Ref.~\cite{BHK99} through a rather technical definition
specific to an effective field theory.
Efimov's Eq.~(\ref{B3-Efimov}) together with the explicit parametrization
of $\Delta(\xi)$ given below in Eqs.~(\ref{expol1}-\ref{expol3})
provides an equivalent definition of $\Lambda_*$.
Note that we measure $B_3$ from the 3-atom threshold,
so $B_3 > B_2$ for $a > 0$. If the universal
function $\Delta(\xi)$ is known, the Efimov binding energies $B_3$
can be calculated as a function of $a$ and $\Lambda_*$
by solving Eq.~(\ref{B3-Efimov}) for different values of the integer $n$.
Eq.~(\ref{B3-Efimov}) has an exact discrete scaling symmetry: if there is an
Efimov state with binding energy $B_3$ for the parameters $a$
and $\Lambda_*$, then there is also an Efimov state with binding energy
$\lambda^2 B_3$ for the parameters $\lambda^{-1} a$ and $\Lambda_*$
if $\lambda=\exp[n'\pi/s_0]$ with $n'$ an integer. 
Because of this symmetry, Eq.~(\ref{B3-Efimov}) defines $\Lambda_*$
only up to multiplicative factors of $\exp[\pi/s_0]$.
If $a>0$, the scattering length $a$ in Eqs.~(\ref{Hxi-def}, 
\ref{B3-Efimov}) can be replaced by $a_B$ defined in Eq.~(\ref{eq:ab}).
The change in the predictions for $B_3$ when $a_B$ is used instead of
$a$ can be taken as an estimate of
the theoretical error associated with effective range corrections.

The universal function $\Delta(\xi)$ could be determined by
solving the 3-body Schr{\"o}dinger equation for the Efimov
binding energies in various potentials whose scattering lengths
are so large that effective range corrections are negligible. It
can be calculated more easily by using the effective field theory
of Ref.~\cite{BHK99} in which the effective range can be set to zero.
In Ref.~\cite{BHK99}, the dependence of the binding energy
on $a$ and $\Lambda_*$ was calculated for the shallowest Efimov state
and $a>0$. In Ref.~\cite{BHK02}, the binding energies of the three lowest
Efimov states were calculated for both signs of $a$ and used
to extract the universal function $\Delta(\xi)$.
%%%%%%%%%%%%%%%%%%%%%%%%%%%%%%%%%%%%%%%%%%%%%%%%%%%%%%%%%%
\begin{figure}[tb]
\begin{center}
%\centerline{\includegraphics[width=12cm,angle=0,clip=true]{efipmhe4.ps}}
\centerline{\includegraphics[width=12cm,angle=0,clip=true]{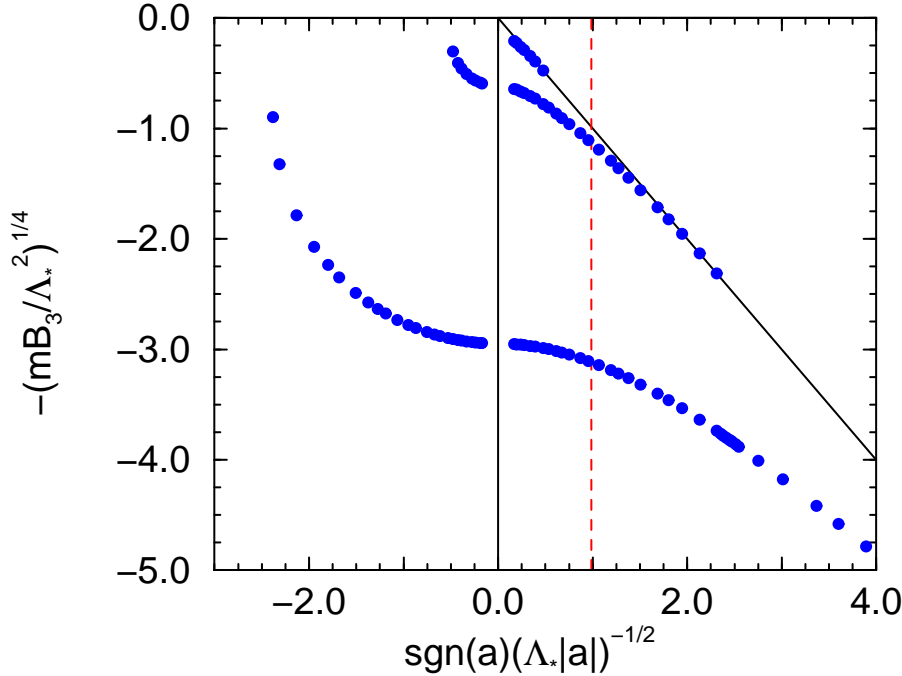}}
\end{center}
\vspace*{-18pt}
\caption{The energy variable $-(mB_3/\hbar^2\Lambda_*^2)^{1/4}$ for three
 shallow Efimov states as a function of ${\rm sgn }(a)(\Lambda_* |a|)^{-1/2}$.
 The vertical dashed line corresponds to the LM2M2 and TTY potentials 
 for $^4$He.}
\label{fig:B3n}
\end{figure}
%%%%%%%%%%%%%%%%%%%%%%%%%%%%%%%%%%%%%%%%%%%%%%%%%%%%%%%%%%%%%%    
In Fig.~\ref{fig:B3n},
we plot $-(mB_3/\hbar^2\Lambda_*^2)^{1/4}$ as a function of
${\rm sgn }(a)(\Lambda_* |a|)^{-1/2}$ for these three
branches of Efimov states.
The binding energies for deeper Efimov states and for shallower states
near $(\Lambda_* |a|)^{-1/2}=0$ can be obtained from the discrete
scaling symmetry. Parametrizations of $\Delta(\xi)$ in various regions 
for $\xi$ were obtained by fitting the Efimov spectrum \cite{BHK02}: 
\bqa
\label{expol1}
&&\mqo2\mqo2\mqo2\xi \in {\textstyle [-{3\pi \over 8},-{\pi \over 4}]:}\,\;
\Delta=3.10x^2 -9.63x -2.18 ,\;\quad x=(-\pi/4-\xi)^{1/2}\,,\\[2pt]
\label{expol2}
&&\mqo2\mqo2\mqo2\xi \in {\textstyle [-{5\pi \over 8},-{3\pi \over 8}]:}\,\;
\Delta=1.17y^3+1.97y^2+2.12y-8.22,\;\quad y=\pi/2+\xi \,,\\[2pt]
\label{expol3}
&&\mqo2\mqo2\mqo2\xi \in {\textstyle [-\pi,-{5\pi \over 8}]:}\,\;
\Delta=0.25z^2+0.28z-9.11,\;\quad z=(\pi+\xi)^2 \exp[-1/(\pi+\xi)^2]\,.
\eqa
These parametrizations deviate from the numerical results by less than 
0.013. The discontinuity at $\xi=-{3\pi\over 8}$ and 
$\xi=-{5\pi\over 8}$ is less than 0.016.
Using Eq.~(\ref{B3-Efimov}) and the parametrizations
(\ref{expol1}-\ref{expol3}), 
the full spectrum of Efimov states can be calculated
as a function of $a$ and $\Lambda_*$. Eq.~(\ref{B3-Efimov})
can also be used as an operational definition of the three-body
parameter $\Lambda_*$, which was originally defined in the
framework of effective field theory \cite{BHK99}. If the 
binding energy $B_3$ of an Efimov state is known either from 
experiment or by solving the three-body Schr{\"o}dinger equation,
we can determine $\Lambda_*$ by demanding that $B_3$ be a 
solution to Eq.~(\ref{B3-Efimov}) for some integer $n$.

A given 2-body potential is characterized by values of $a$ and
$\Lambda_*$ and corresponds to a vertical line in Fig.~\ref{fig:B3n}.
The dashed line shown corresponds to the LM2M2 and TTY potentials for $^4$He 
atoms. The intersections of this line with the binding energy curves
correspond to the infinitely many Efimov states. The two intersections 
visible in the figure correspond to the excited state and the ground state
of the $^4$He trimer. The third bound state predicted by Efimov's
equation has a binding energy that is approximately $515\, B_3^{(0)}
\approx 67$ K. This is much larger than the natural low-energy scale 
$\hbar^2/ml^2$ which is 0.4 K. 
This state and all the deeper Efimov states are therefore artifacts 
of the limit $a \gg l$.

We can use one of the trimer binding energies as the input to
determine $\Lambda_*$. It is safer to use the binding energy $B_3^{(1)}$
of the excited state, because it is least affected by the high-energy effects 
that cut off the Efimov spectrum.
The most accurate calculations of $B_3^{(1)}$
have been obtained by solving the Faddeev equations in
the hyperspherical representation \cite{NFJ98},
in configuration space \cite{RoY00},
and with hard-core boundary conditions \cite{MSSK01}.
These methods give results that agree to within 0.6\%.
The results of Ref.~\cite{MSSK01} for $B_3^{(0)}$ and $B_3^{(1)}$ for 
the HFDHE2, HFD-B, LM2M2, and TTY potentials are given in Table~\ref{tab1}.
%%%%%%%%%%%%%%%%%%%%%%%%%%%%%%%%%%%%%%%%%%%%%%%%%%%%%%%%%%%%%%%%%%%%%%%%
\begin{table}[htb]
\begin{tabular}{c||c|c||c|c|c||c|c|c}
Potential & $B_3^{(0)}$ & $B_3^{(1)}$ &
$\; a_B \Lambda_*\;$ & $B_3^{(0)}$ (LO) & $B_3^{(0)}$ (NLO) &
$\; a\Lambda_*\;$ & $B_3^{(0)}$ (LO) & $B_3^{(0)}$ (NLO)\\
\hline\hline
HFDHE2 & 116.7 & 1.67 & 1.258 & 118.5 & 116.7 & 1.364 & 129.1 & 119.3\\
HFD-B  & 132.5 & 2.74 & 0.922 & 137.5 & [132.5] & 1.051 & 159.7 & [132.5]\\
LM2M2  & 125.9 & 2.28 & 1.033 & 130.3 & 126.8 & 1.155 & 147.4 & 128.6\\
TTY    & 125.8 & 2.28 & 1.025 & 129.1 & 125.6 & 1.147 & 146.4 & 127.5
\end{tabular}
\caption{The trimer binding energies $B_3^{(0)}$ and $B_3^{(1)}$
in mK measured from the three-atom threshold for four $^4$He potentials.
The calculated values from Ref.~\protect\cite{MSSK01} are in
columns 1--2. The values of $a\Lambda_*$, the universality predictions 
for $B_3^{(0)}$, and the predictions for $B_3^{(0)}$ including effective
range corrections using $B_2$ and $B_3^{(1)}$ as input are in columns 
3--5. The corresponding values using $a$ and $B_3^{(1)}$ as input
are in columns 6--8. Numbers in brackets were used as input.
}
\label{tab1}
\end{table}
%%%%%%%%%%%%%%%%%%%%%%%%%%%%%%%%%%%%%%%%%%%%%%%%%%%%%%%%%%%%%%%%%%%%%%%%  
The results for the binding energy $B_3^{(0)}$ of the ground state
of the trimer agree well with Diffusion Monte Carlo calculations
\cite{Lew97} which give $B_3^{(0)}=(131.0 \pm 0.7)$ mK for the HFD-B
potential and $(125.5 \pm 0.6)$ mK for the TTY potential.
Taking the calculated dimer binding energy $B_2$ as the two-body 
input, we determine $\Lambda_*$ by demanding that $B_3^{(1)}$ satisfy
(\ref{B3-Efimov}) with $n=1$. Solving Eq.~(\ref{B3-Efimov}) with $n=2$,
we obtain the predictions for $B_3^{(0)}$ in
column 4 of Table~\ref{tab1}. The predictions 
are only 1-4\% higher than the calculated values.
If we use the calculated values 
of $a$ as input instead of $B_2$, we obtain the predicted values of
$B_3^{(0)}$ in column 7 of Table~\ref{tab1}. These values
are larger than the calculated ones by 11-21\%.
The difference between these predictions and those obtained by
using $B_2$ and $B_3^{(1)}$ as the input gives an indication
of the size of effective range corrections.
The predictions are labelled LO in Table~\ref{tab1}, because they 
are the universality prediction at \lq\lq leading order'' in $r_s/a$.

The numerical values of $\Lambda_*$ for the four potentials in the 
Table~\ref{tab1} are nearly the same.  If we use $a_B$ and $B_3^{(0)}$
as the input, we obtain  $\Lambda_*= 0.0107$ \AA$^{-1}$
for the LM2M2 and TTY potentials.  The values for the other two 
potentials differ by less than 3\%. If we use $a$ and $B_3^{(0)}$
as the input, we obtain  $\Lambda_*= 0.0115$ \AA$^{-1}$
for the LM2M2 and TTY potentials.  The values for the other two 
potentials differ by less than 5\%.
The small differences between the values of $\Lambda_*$ for these 
potentials illustrates the fact that $\Lambda_*$ tends to be insensitive 
to the parameter in the potential that is tuned to make the 
scattering length large.

\begin{figure}[tb]
\begin{center}
%\centerline{\includegraphics[width=12cm,angle=0,clip=true]{he4b3phil.ps}}
\centerline{\includegraphics[width=12cm,angle=0,clip=true]{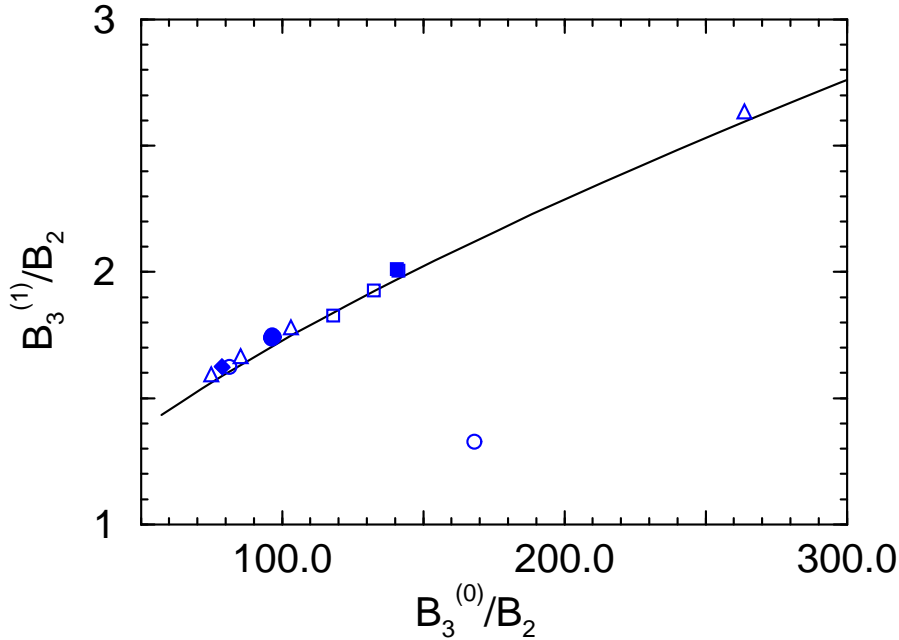}}
\end{center}
\vspace*{-18pt}
\caption{The excited state energy $B_3^{(1)}$ as a function of the ground 
 state energy $B_3^{(0)}$. The solid line is the universal scaling curve
 predicted by Eq.~(\ref{B3-Efimov}). The filled symbols show the results
 from Motovilov et al. \protect\cite{MSSK01},
 while the open symbols display the results from various 
 other calculations \protect\cite{NFJ98,RoY00,ELG96,CoG86,Gon99}. 
 The results for the LM2M2/TTY, HFDHE2, 
 and HFD-B potentials are indicated by the circles, squares, and 
 diamonds, respectively. The open triangles show results from
 Ref.~\protect\cite{RoY00} for four other potentials.}
\label{fig:he4b3phil}
\end{figure}

The availability of accurate calculations of $B_3^{(0)}$  and 
$B_3^{(1)}$ for various older $^4$He potentials can be used to demonstrate
the nontrivial nature of universality in the three-body sector.
Different potentials that give
a large two-body scattering length should correspond to different
values of $\Lambda_*$. The scaling variables
$B_3^{(0)}/B_2$  and  $B_3^{(1)}/B_2$ are functions of $a\Lambda_*$
only. If we eliminate $\Lambda_*$, we obtain a prediction 
for $B_3^{(1)}/B_2$ as a universal function of $B_3^{(0)}/B_2$.
A closely related scaling function that expresses $B_3^{(1)}/
B_3^{(0)}$ as a function of $B_2/B_3^{(0)}$
has been calculated by Frederico, 
Tomio, Delfino, and Amorim using the renormalized zero-range
model \cite{FTD99,DFT00}. We have reproduced their scaling function
using the solution to Efimov's equation (\ref{B3-Efimov}). In
Fig.~\ref{fig:he4b3phil}, our calculation of
the universal scaling function relating
$B_3^{(1)}/B_2$ to $B_3^{(0)}/B_2$ is shown as a solid line. As $\Lambda_*$
increases, one moves along the solid line to the right. The data
points in Fig.~\ref{fig:he4b3phil} are the results from calculations
with various $^4$He potentials.
%%%%%%%%%%%%%%%%%%%%%%%%%%%%%%%%%%%%%%%%%%%%%%%%%%%%%%%%%%
%%%%%%%%%%%%%%%%%%%%%%%%%%%%%%%%%%%%%%%%%%%%%%%%%%%%%%%%%%%%%%
The filled symbols show the results
from Motovilov et al. \protect\cite{MSSK01} which we used
to determine $\Lambda_*$ for each potential, 
while the open symbols display the 
results from various other calculations 
\protect\cite{NFJ98,RoY00,ELG96,CoG86,Gon99}. 
The results for the LM2M2/TTY, HFDHE2, 
and HFD-B potentials are displayed by the circles, squares, and 
diamonds, respectively.
The open triangles show results from
Ref.~\protect\cite{RoY00} for four other potentials.

The points fall very close to the universal scaling
curve, with the exception of the result of Ref.~\cite{Gon99} for the 
LM2M2 potential, which lies well below. In that paper,
the overall strength of the potential was varied such that 
$B_2$ passed through zero. The results failed to exhibit the
Efimov effect of an accumulation of three-body bound states at 
threshold as $B_2 \to \infty$. The numerical accuracy of this
calculation has been questioned in Ref.~\cite{ELG01}.
All the remaining results fall along the
universal scaling curve. This demonstrates that the dominant effect
of the different potentials on the trimer binding energies can be 
described by a single parameter which we can identify with $\Lambda_*$.
 
It is interesting to note that
calculations using only the lowest adiabatic hyperspherical potential
\cite{ELG96} or the lowest orbital angular momentum wave function give
results that already lie near the scaling curve. Including additional
adiabatic potentials or higher orbital angular momenta $L$ moves the
point to the right along the scaling curve until convergence is reached.
This trend is most evident in the calculation of Ref.~\cite{MSSK01}
where the partial results for $L_{max}=0,2$ can be compared with the 
fully converged result with $L_{max}=4$ (see Table 2 and 3 
in Ref.~\cite{MSSK01}). 

Note that the most accurate points in Fig.~\ref{fig:he4b3phil} all
lie systematically above the scaling curve by approximately the same 
amount. One can interpret this deviation as being due to effective
range effects.
These effects are included in potential models, but the effective
range is set to zero in the renormalized zero range model and in
the effective field theory that were used to calculate the scaling curve.
It should be possible to account for these differences quantitatively
by taking into account effective range corrections \cite{Efi91,HaM01}.
Since $r_s/a\approx 0.07$, we expect that including the effective 
range corrections as a first order perturbation would shift the
scaling curve by a small amount, bringing it into
better agreement with the calculated points.

We can take into account the effective range corrections 
to $B_3^{(0)}$ approximately if we assume that the deviation
$\Delta B_3^{(0)}$ of the leading order universality prediction
from the calculated value comes almost entirely from a correction linear
in $r_s$. The calculated result for one potential can then be used 
to estimate the effective range corrections for the 
others. Choosing $B_3^{(0)}$ for the HFD-B potential as the input and
denoting the deviations of $B_3^{(0)}$ from the 
leading order universality predictions
by $\Delta B_3^{(0)}$, we can estimate the effective range corrections
for any other potential by
\beq
\left( \frac{\Delta B_3^{(0)}}{B_2}\right)_{\rm pot}=
\left( \frac{\Delta B_3^{(0)}}{B_2}\right)_{\rm HFD-B}\;
\frac{(r_s/a_B)_{\rm pot}}{(r_s/a_B)_{\rm HFD-B}}\,.
\eeq
Since the various $^4$He potentials have similar values of $r_s$, the
shift $\Delta B_3^{(0)}$ is almost the same for all the potentials.
The resulting predictions for $B_3^{(0)}$ are shown in column 5
of Table~\ref{tab1}. The corresponding prediction using $a$ and
$B_3^{(1)}$ as the input are shown in column 8.
The predictions are labelled NLO in Table~\ref{tab1}, because they are
approximate universality predictions at \lq\lq next-to-leading order''
in $r_s/a$. For each of the HFDHE2, LM2M2, and TTY potentials,
the two NLO predictions differ by less than 3\%. They also
differ from the calculated results in column 1 by less than 3\%.

We can use the results in Table~\ref{tab1} to give universality 
predictions for $B_3^{(0)}$ for potentials other than HFD-B both at 
leading order (LO) and next-to-leading order (NLO) in the effective
range. As our best estimate, we take the average of the two
predictions obtained by using $B_2$ and $a$ as the two-body input.
We take the difference to be an estimate of the theoretical
error. The universality predictions for the TTY potential are
\bqa
\mbox{LO}:&& \quad B_3^{(0)}=138 \pm 17\mbox{ mK}
\qquad\qquad \mbox{(TTY)}\,,\nonumber\\
\mbox{NLO}:&& \quad B_3^{(0)}=127 \pm 2\mbox{ mK}
\qquad\qquad \;\mbox{\ (TTY)}\,.
\label{eq:b30pred}
\eqa
The calculated value in column 1 of Table~\ref{tab1} lies within the 
error bar for both predictions. Note that including effective range 
corrections decreases the size of the error bar by an order of magnitude.

The identification of the excited state of the $^4$He trimer as an Efimov 
state is well established
\cite{LDD77,NFJ98,MSSK01,RoY00,ELG96,BHK99,BHK02,FTD99}.
We now discuss the question of whether the ground state of the $^4$He 
trimer should be identified as an Efimov state.
The good agreement between the universality prediction for
$B_3^{(0)}$ and the calculated value could be fortuitous. Some authors 
have used as the criterion for an Efimov state that a sufficiently 
large increase in the strength of the two-body potential should make it
unstable to decay into an atom and a dimer. Increasing the strength 
of the two-body potential decreases the scattering length. This tends
to move the vertical dashed line in Fig.~\ref{fig:B3n} to the right.
A sufficiently large shift in the vertical line will move it beyond the
point where that branch of Efimov states terminates on the line 
corresponding to the dimer binding energy. According to this criterion,
the excited state of the trimer is an Efimov state but the ground state
is not. However, we argue that the criterion for an Efimov state
should not be based on how its binding energy behaves under a
large deformation of the strength of the two-body potential, but
on how it behaves under arbitrary small deformations of the potential.
If it is an Efimov state, any small deformation of the two-body 
potential will move its binding energy along the universal scaling
curve in Fig.~\ref{fig:he4b3phil}. The various model potentials
for $^4$He can be interpreted as deformations of the \lq\lq true''
$^4$He potential.\footnote{The short-range part of the 
\lq\lq true'' potential can be defined by the leading order Born-Oppenheimer
approximation. Corrections to this approximation are suppressed by 
$m_e/m$, where $m_e$ is the electron mass.}
The fact that the binding energies for these potentials
lie along the universal curve is convincing evidence that the ground 
state of the $^4$He trimer is  an Efimov state.

If the true binding energy $B_3^{(1)}$ of the excited $^4$He trimer
was measured and found to
disagree with the calculated value using modern potentials, it 
would indicate that those potentials are not sufficiently accurate to
predict low-energy three-body observables.
Using the universality approach, there would be no need to improve
the potential in order to predict these observables. We could
simply take the measured value of $B_3^{(1)}$ as the input required
to determine $\Lambda_*$. If we did choose to improve the
potential, universality implies that to get predictions for low-energy 
three-body observables with errors of order $r_s/a$, it would be sufficient 
to introduce a two-parameter deformation of the 
short-distance part of the potential and tune
both parameters simultaneously so that the potential gives the correct
values for $B_2$ and $B_3^{(1)}$. Alternatively, we could leave 
the two-body potential unchanged, but instead introduce an
artificial short-range  three-body
potential and tune its strength in order to get the correct value
for $B_3^{(1)}$. This is essentially what is done in the effective
field theory approach; the parameter $\Lambda_*$ is varied by adjusting
the strength of a three-body contact interaction \cite{BHK99}. 
In the case of $^4$He atoms, the \lq\lq true'' three-body potential
decreases the binding energy $B_3^{(0)}$ of the ground state
trimer by about 0.3 mK \cite{Lew97}. Its effect on $B_3^{(1)}$
should be much smaller, because the excited state is much larger in
size. Thus the effects of the \lq\lq true'' three-body potential on 
low-energy three-body observables should be very small.
However, universality implies that the dominant
effect on low-energy three-body observables from a deformation of
the two-body potential that leaves the scattering length fixed is 
equivalent to the effect of adding a three-body potential.

\section{Atom-Dimer Scattering}
\label{sec:2plus1}

The differential cross section for the elastic scattering of an 
atom and a dimer with wave numbers $k$ in the center-of-mass frame
has the form 
\beq
\frac{d\sigma}{d\Omega}=\left|\sum_{L=0}^\infty
\frac{2L+1}{k\cot\delta_L(k)-ik} P_L(\cos\theta)\right|^2\,,
\eeq
where $\delta_L(k)$ is the phase shift for the $L^{\rm th}$ partial
wave. These phase shifts are real-valued below the dimer
breakup threshold at $k=(4mB_2/3\hbar^2)^{1/2}$. Above that threshold, they 
become complex-valued because of the inelasticity from scattering 
into three-atom final states. 

If the two-body scattering length is large, 
the cross section for low energies $E\simle \hbar^2/ma^2$
has a universal form. For $L \geq 1$, the phase shifts $\delta_L(k)$
are universal functions of $ka$ only. To the best of our knowledge,
they have not been calculated. The $L=0$ phase shift $\delta_0(k)$
is also universal, but it depends on $a\Lambda_*$ as well as on $ka$.
The general structure of the dependence on $a\Lambda_*$ was deduced 
by Efimov \cite{Efi71}. For $k$ below the breakup threshold,
$ka\cot\delta_0$ can be written as
\beq
ka \cot\delta_0 = c_1(ka)+c_2(ka)\cot[s_0\ln(a\Lambda_*)+\phi(ka)]\,,
\qquad 0\leq ka \leq \frac{2}{\sqrt{3}}\,,
\label{kcot-Efimov}
\eeq
where $c_1(ka)$, $c_2(ka)$, and $\phi(ka)$ are unknown
universal functions that satisfy the constraints
$c_1(2/\sqrt{3})=0$ and $c_2(2/\sqrt{3})=2/\sqrt{3}$ \cite{review}.
Using the effective field theory of Ref.~\cite{BHK99}, we have 
calculated these functions. The results can be parametrized as follows:
\bqa
c_1(ka)&=&-0.22+0.39\,k^2 a^2 -0.17\,k^4 a^4\,,\nonumber\\
c_2(ka)&=&0.32+0.82\,k^2 a^2 -0.14\,k^4 a^4\,,\nonumber\\
\phi(ka)&=&2.64-0.83\,k^2 a^2 +0.23\,k^4 a^4\,. 
\label{kcot-par}
\eqa
The atom-dimer scattering length $a_{12}$ and effective range
$r_{s,12}$ are defined by the low-energy limit of the S-wave phase shift
by an equation analogous to (\ref{eq:ere}).
From Eqs.~(\ref{kcot-Efimov}, \ref{kcot-par}), we obtain after the use
of trigonometric identities
\bqa
\label{eq:a-par}
a_{12}&=&a\left( 1.46 - 2.15 \tan[s_0 \ln(a\Lambda_*)+ 0.09 ]\right)
\,,  \\
\label{eq:re-par}
r_{s,12}&=&a\left( 1.30 - 1.64 \tan[s_0 \ln(a\Lambda_*)+ 1.07 ]
+ 0.53 \tan^2[s_0 \ln(a\Lambda_*)+ 1.07 ]\right)\,.
\eqa

\begin{figure}[tb]
\begin{center}
%\centerline{\includegraphics[width=15cm,angle=0,clip=true]{are1.ps}}
\centerline{\includegraphics[width=15cm,angle=0,clip=true]{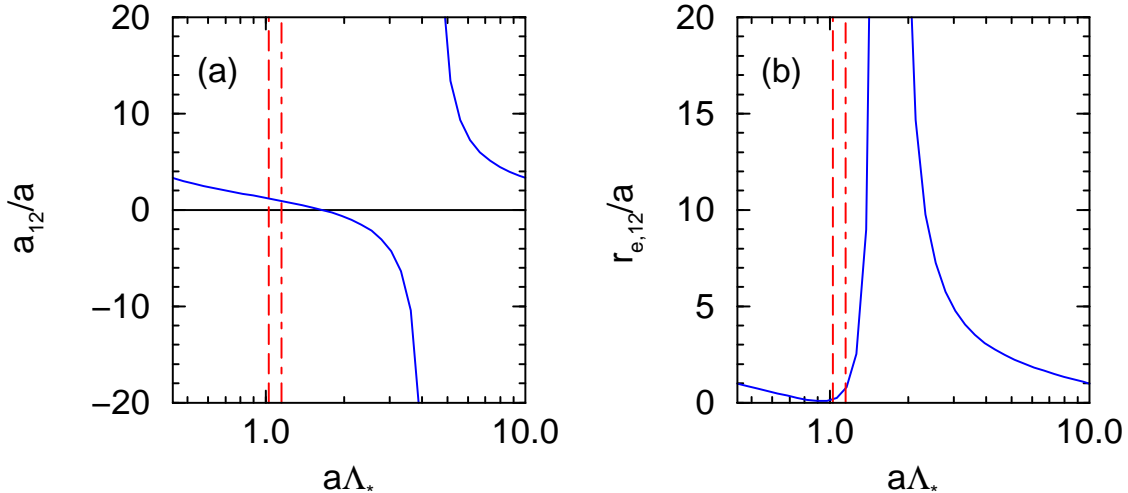}}
\end{center}
\vspace*{-18pt}
\caption{The atom-dimer scattering length $a_{12}/a$ [(a)]
and effective range $r_{s,12}/a$ [(b)] as a function of $a\Lambda_*$.
The values of $a_B\Lambda_*$ and $a\Lambda_*$ for the TTY
potential are indicated by the vertical dashed and dot-dashed lines,
respectively.}
\label{fig:are1}
\end{figure}

The atom-dimer scattering lengths $a_{12}$ for the 
HFD-B, LM2M2, and TTY potentials were calculated in Ref.~\cite{MSSK01}, 
and the results are given in column 1 of Table~\ref{tab2}.
Using the values of $\Lambda_*$ determined in the previous section,
we can predict the atom-dimer scattering length and compare 
with the calculated values. The leading order universality predictions
for $a_{12}$ and $r_{s,12}$  are given in columns 3, 5, 7, and 9
of Table~\ref{tab2}.
%%%%%%%%%%%%%%%%%%%%%%%%%%%%%%%%%%%%%%%%%%%%%%%%%%%%%%%%%%%%%%%%%%%%%%%%
\begin{table}[htb]
\begin{tabular}{c||c||c|c|c|c||c|c|c|c}
Potential & $a_{12}$ &
$\; a_B\Lambda_*\;$ & $a_{12}$ (LO) & $a_{12}$ (NLO) & $r_{s,12}$ (LO) & 
$\; a\Lambda_*\;$ & $a_{12}$ (LO) & $a_{12}$ (NLO) & $r_{s,12}$ (LO)\\
\hline\hline
HFDHE2 & $-$  &  1.258 & 87.9  & 103(5) & 278 & 1.364 & 65.8 & 101(5) & 902\\
HFD-B  & 135(5) & 0.922 & 120.2 & [135(5)] & 6.4 & 1.051 & 100.4 & [135(5)] 
& 18.6\\
LM2M2  & 131(5) & 1.033 & 113.1 & 128(5) & 16.0 & 1.155 & 92.8 & 128(5) & 75\\
TTY    & 131(5) & 1.025 & 114.5 & 129(5) & 14.4 & 1.147 & 94.0 & 129(5) & 69
\end{tabular}
\caption{The atom-dimer scattering lengths $a_{12}$ 
and effective ranges $r_{s,12}$ in \AA\ for four $^4$He potentials.
The calculated values of $a_{12}$ from Ref.~\protect\cite{MSSK01}
are in column 1. The values of $a\Lambda_*$, the universality predictions
for $a_{12}$, the predictions for $a_{12}$ including effective
range corrections, and the universality predictions for $r_{s,12}$
using $B_2$ and  $B_3^{(1)}$ as input are in columns 2--5.
The corresponding predictions using $a$ and $B_3^{(1)}$ as input are in 
columns 6--9. Numbers in brackets were used as input.}
\label{tab2}
\end{table}
%%%%%%%%%%%%%%%%%%%%%%%%%%%%%%%%%%%%%%%%%%%%%%%%%%%%%%%%%%%%%%%%%%%%%%%%
If $B_2$ and $B_3^{(1)}$ are used as inputs, the predictions for $a_{12}$
are smaller than the calculated values by about 13\%. If $a$ and $B_3^{(1)}$
are used as inputs, the predictions are smaller than the calculated values 
by about 28\%. Note that the predictions for $r_{s,12}$ differ by as much as
a factor of five depending on whether $B_2$ or $a$ is used as the
two-body input.
In Fig~\ref{fig:are1}, we show the atom-dimer scattering parameters
$a_{12}$ and $r_{s,12}$ as functions of $a\Lambda_*$.
%%%%%%%%%%%%%%%%%%%%%%%%%%%%%%%%%%%%%%%%%%%%%%%%%%%%%%%%%%
%%%%%%%%%%%%%%%%%%%%%%%%%%%%%%%%%%%%%%%%%%%%%%%%%%%%%%%%%%%%%% 
The values of $a_B\Lambda_*$ and $a\Lambda_*$ for the TTY
potential are indicated by the vertical dashed and dot-dashed lines,
respectively. Note that $r_{s,12}$ is positive definite.
It achieves a minimum value that is very close to zero
near $a\Lambda_*=0.94$ and diverges at $a\Lambda_*=1.64$.
The extracted values of $a\Lambda_*$ for $^4$He are fortuitously 
in the interval between the minimum and the divergence
where the effective range changes rapidly with $a\Lambda_*$.
This leads to a large difference in the values of $r_{s,12}$ 
obtained from using $a$ or $B_2$ as the two-body input.

\begin{figure}[tb]
\begin{center}
%\centerline{\includegraphics[width=12cm,angle=0,clip=true]{TTYecmb2.ps}}
\centerline{\includegraphics[width=12cm,angle=0,clip=true]{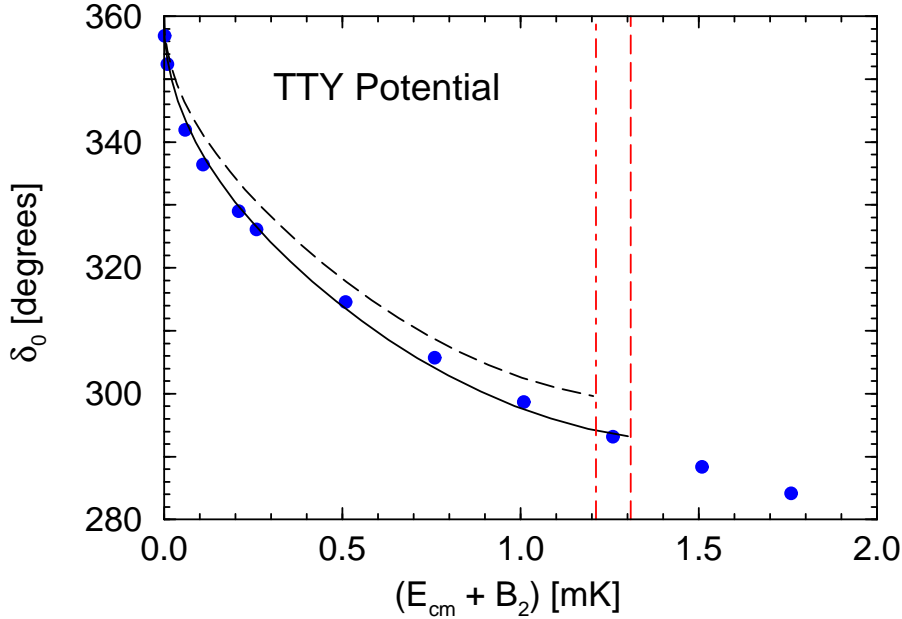}}
\end{center}
\vspace*{-18pt}
\caption{The S-wave scattering phase shifts $\delta_0$ for the TTY 
potential using $B_2$ as the two-body input (solid line) and using
$a$ as the two-body input (dashed line) as a function of the 
center-of-mass energy (with the scattering threshold defined
as zero energy). The filled circles show the
fully converged calculation of Ref.~\protect\cite{MSSK01} with 
$L_{max}=4$. The vertical dashed and dot-dashed lines indicate the 
dimer breakup threshold for $B_2$ and $a$ as the two-body input,
respectively.}
\label{fig:he4phase}
\end{figure}

In Fig.~\ref{fig:he4phase}, we show the S-wave scattering phase shift 
$\delta_0(k)$ for the TTY potential as a function of the center-of-mass 
energy $E_{cm}$ shifted by the dimer binding energy so that the scattering
threshold is at zero energy,
\beq
E_{cm}+B_2=\frac{3 \hbar^2 k^2}{4 m}\,,
\eeq
where $k$ is the wave number of the atom or the dimer.
This shifted energy variable has the advantage that the
position of the scattering threshold is independent of whether $B_2$ or $a$ 
is taken as the two-body input.
%%%%%%%%%%%%%%%%%%%%%%%%%%%%%%%%%%%%%%%%%%%%%%%%%%%%%%%%%%
%%%%%%%%%%%%%%%%%%%%%%%%%%%%%%%%%%%%%%%%%%%%%%%%%%%%%%%%%%%%%% 
The solid and dashed lines show the universality prediction with $B_2$
and $a$ as the two-body input, respectively. The vertical dashed
and dot-dashed lines indicate the dimer breakup threshold for $B_2$
and $a$ as the two-body input, respectively.
The filled circles show the
results of Ref.~\protect\cite{MSSK01}, which were obtained by solving
the Faddeev equation with hard-core boundary conditions. The results are
in good agreement with the error band defined by the solid and dashed curves.

\begin{figure}[tb]
\begin{center}
%\centerline{\includegraphics[width=15cm,angle=0,clip=true]{he4philipall.ps}}
\centerline{\includegraphics[width=15cm,angle=0,clip=true]{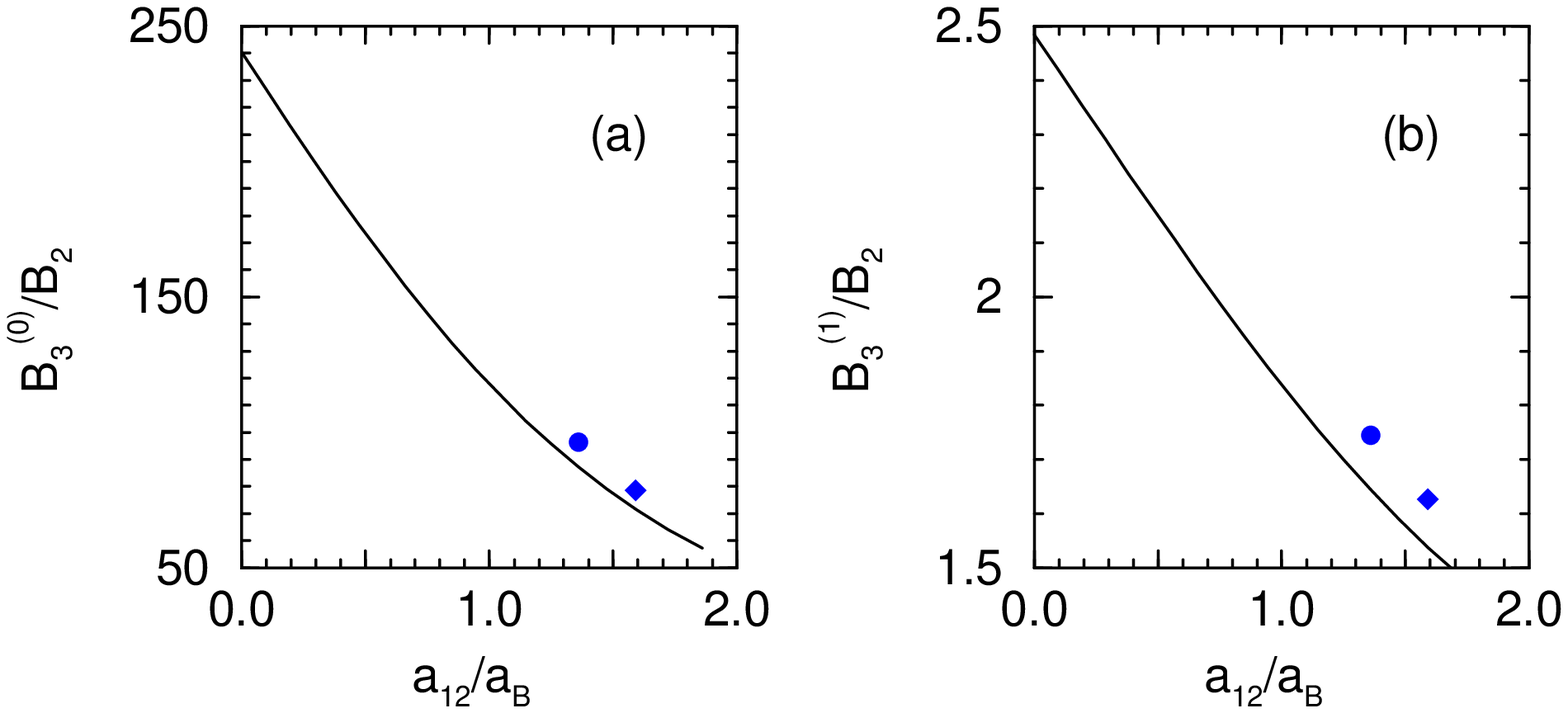}}
\end{center}
\vspace*{-18pt}
\caption{The Phillips line for (a) the trimer ground state and (b)
excited state. The solid line gives the universality prediction,
while the data points show the results of Ref.~\protect\cite{MSSK01}
for the HFD-B (diamonds) and LM2M2/TTY potentials (circles).}
\label{fig:he4phil}
\end{figure}

Universality implies that the scaling variable $a_{12}/a_B$ is a universal
function of $a\Lambda_*$. By eliminating $\Lambda_*$, we can express 
$B_3^{(n)}/B_2$ as a universal function of $a_{12}/a_B$. The various
$^4$He potentials should all give binding energies and scattering lengths 
that lie along this curve. In nuclear physics, a similar correlation has 
been observed between the spin-doublet neutron-deuteron scattering length
$a_{12}$ and the binding energy of the triton $B_3$. Calculations
of $B_3$ and $a_{12}$ using various potential models for the 
nucleon-nucleon interaction give results that cluster along a line in the
$a_{12}$-$B_3$ plane called the Phillips line \cite{Phi68}. The 
observed values of $B_3$ and $a_{12}$ also lie on that line.
Modern nucleon-nucleon potentials predict a value for $B_3$ that is
about 5-10\% below the measured value.
Accurate values for both $a_{12}$ and $B_3$ are obtained by adding a 
short-range three-body potential and adjusting one parameter to
reproduce the measured triton binding energy. The Phillips line
in nuclear physics arises from the large S-wave scattering length 
in both the spin-triplet ($r_s/a\approx 1/3$)
and spin-singlet ($r_s/|a|\approx 1/8$) nucleon-nucleon channels 
\cite{Efi88,BHK00}. In the case of $^4$He, there are
two Phillips lines: one for the $^4$He trimer ground state and 
one for the excited state. These Phillips lines are shown 
in Figs.~\ref{fig:he4phil}(a) and \ref{fig:he4phil}(b), respectively.
%%%%%%%%%%%%%%%%%%%%%%%%%%%%%%%%%%%%%%%%%%%%%%%%%%%%%%%%%%
%%%%%%%%%%%%%%%%%%%%%%%%%%%%%%%%%%%%%%%%%%%%%%%%%%%%%%%%%%%%%% 
The solid line is the universality prediction from 
Eqs.~(\ref{B3-Efimov}) and (\ref{eq:a-par}).
As $\Lambda_*$ increases, one moves along the solid line to the left.
The data points show the results of Ref.~\protect\cite{MSSK01}
for the LM2M2/TTY (circles) and HFD-B potentials (diamonds). 
They lie close to the Phillips lines as expected from universality.
For both potentials, the points lie slightly above the scaling curves, 
consistent with small effective range corrections.

We can use the calculated result for $a_{12}$ for the HFD-B potential
from Ref.~\cite{MSSK01} to estimate the effective range corrections for the
other potentials. Denoting the deviation of $a_{12}$ from the universality
prediction by $\Delta a_{12}$, we can estimate the effective range 
correction by
\beq
\left( \frac{\Delta a_{12}}{a_B}\right)_{\rm pot}=
\left( \frac{\Delta a_{12}}{a_B}\right)_{\rm TTY}\;
\frac{(r_s/a_B)_{\rm pot}}{(r_s/a_B)_{\rm TTY}}\,.
\eeq
The resulting predictions for $a_{12}$ are shown in column 4
of Table~\ref{tab2}. The corresponding predictions using $a$ and
$B_3^{(1)}$ as the input are shown in column 8.
The two NLO predictions agree to within 2\%.
For the LM2M2 and TTY potentials, they agree with the 
calculated results in column 1 to within the error bars.

We can use the results in Table~\ref{tab2} to give universality 
predictions for potentials other than HFD-B for $a_{12}$ both
at leading order and next-to-leading order in the effective range.
As the prediction and the theoretical error in
$a_{12}$, we take the average and the difference of the predictions
obtained by using $B_2$ and $a$ as the two-body input. 
The universality predictions for the TTY potential are
\bqa
\mbox{LO}: &&\quad a_{12} = (104\pm 21)\mbox{ \AA}
\qquad\qquad\mbox{(TTY)}\,,\nonumber \\
\mbox{NLO}:&& \quad a_{12} = (129 \pm 5)\mbox{ \AA}
\qquad\qquad\;\mbox{\ (TTY)}\,.
\label{eq:a12pred}
\eqa
The LO and NLO predictions for the TTY potential both agree to 
within errors with the calculated value in Table~\ref{tab2}. 
Including the effective range corrections decreases the 
error by about an order of magnitude. 
The error in the NLO prediction in (\ref{eq:a12pred}) is dominated
by the error in the calculated value for the HFD-B potential.
There is no accurate calculation for $r_{s,12}$ for any
of the $^4$He potentials. Thus we can only give a leading order
universality prediction for $r_{s,12}$.
Because $r_{s,12}$ is positive definite and because it is
so sensitive to the precise value of $a\Lambda_*$, we take the 
universality prediction for $r_{s,12}$ to be the geometric mean of the 
predictions obtained by using either $B_2$ or $a$ as input. We take the 
theoretical uncertainty to be a multiplicative factor equal to
the ratio of the two predictions. The resulting leading order
universality prediction for the TTY potential is then
\beq
\label{eq:r12pred}
\mbox{LO}:\qquad r_{s,12} = (32^{+121}_{-25})\mbox{ \AA}
\qquad\qquad\mbox{(TTY)}\,.
\eeq
In spite of the large error bars, we can predict with confidence
that $r_{s,12}$ is positive, because the expression (\ref{eq:re-par})
is positive definite.

\section{Three-Body Recombination}
\label{sec:3brec}
Three-body recombination is the process in which two of the 
three incoming atoms form a dimer and the third atom recoils to balance
energy and momentum. The rate of three-body recombination events 
per unit time and unit volume in a gas of cold atoms is proportional 
to the third power of the number density \cite{FRS96}: $\nu =\alpha n^3$.
The recombination rate constant $\alpha$ is a complicated function
of the momenta of the three incoming atoms. At threshold, all three 
momenta vanish and $\alpha$ reduces to a number. The total three-body
recombination rate is the sum of the rates for all the dimers.

If the scattering length $a$ is large and positive, there is a shallow
dimer with $B_2=\hbar^2/ma^2$. The rate constant $\alpha$ for
recombination into the shallow dimer at threshold must be a universal 
function of $a\Lambda_*$. It was calculated in 
Ref.~\cite{BBH00} using the effective field theory of Ref.~\cite{BHK99}.
The result can be parametrized as
\beq
\label{3brec-par}
\alpha = 67.1 \sin^2[s_0\ln(a\Lambda_*)+0.19]
         \,\frac{\hbar a^4}{m}\,\,.
\eeq
This remarkable oscillatory dependence on $\ln(a\Lambda_*)$ was 
previously observed in calculations using the hidden crossing theory
\cite{NiM99} and the adiabatic hyperspherical 
representation \cite{EGB99}.
In the hyperspherical representation, the oscillatory behavior
arises from interference between two pathways from the incoming 
channel on the 2$^{\rm nd}$ adiabatic potential to the outgoing
channel on the 1$^{\rm st}$ adiabatic potential.
Effective field theory allows the argument of the $\sin^2$ to
be determined in terms of the same 
three-body parameter $\Lambda_*$ that enters atom-dimer scattering
and the trimer binding energies. 

\begin{figure}[tb]
\begin{center}
%\centerline{\includegraphics[width=12cm,angle=0,clip=true]{alpha1.ps}}
\centerline{\includegraphics[width=12cm,angle=0,clip=true]{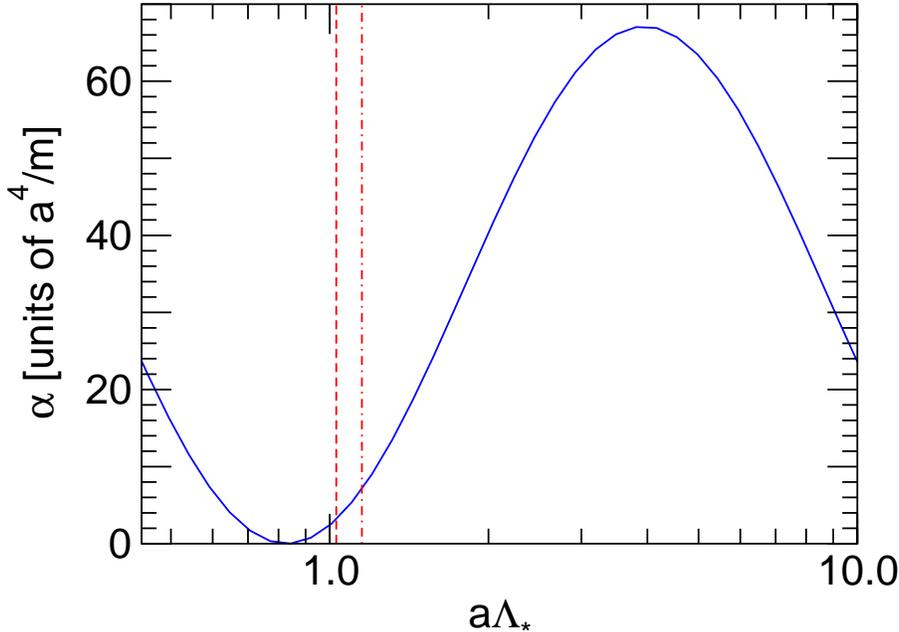}}
\end{center}
\vspace*{-1cm}
\caption{The three-body recombination constant at threshold $\alpha$ in
units of $\hbar a^4/m$ as a function of $a\Lambda_*$ (solid line).
The values of $a_B\Lambda_*$ and $a\Lambda_*$ for the TTY
potential are indicated by the vertical dashed and dot-dashed
lines, respectively.}
\label{fig:alpha}
\end{figure}

Using the values of $\Lambda_*$ determined in Section \ref{sec:bound},
we can predict the three-body recombination constant $\alpha$
from Eq. (\ref{3brec-par}). 
%%%%%%%%%%%%%%%%%%%%%%%%%%%%%%%%%%%%%%%%%%%%%%%%%%%%%%%%%%%%%%%%%%%%%%%%
\begin{table}[htb]
\begin{tabular}{c||c|c||c|c}
Potential &
$\; a_B\Lambda_*\;$ & $\alpha$ (LO)
& $\; a\Lambda_*\;$ & $\alpha$ (LO) \\
\hline\hline
HFDHE2 & 1.258 & $3.79$ & 1.363 & $5.95$ \\
HFD-B  & 0.922 & $0.064$ & 1.051 & $0.37$ \\
LM2M2  & 1.033 & $0.45$ & 1.155 & $1.16$ \\
TTY    & 1.025 & $0.41$ & 1.147 & $1.11$ 
\end{tabular}
\caption{The three-body recombination constant at threshold $\alpha$
in $10^{-27}\,$cm$^6$/s. 
The leading order predictions from universality
using $B_2$ and $B_3^{(1)}$ ($a$ and $B_3^{(1)}$)
as input are in column 2 (4). The corresponding values
of $a\Lambda_*$ are given in column 1 (3). }
\label{tab3}
\end{table}
%%%%%%%%%%%%%%%%%%%%%%%%%%%%%%%%%%%%%%%%%%%%%%%%%%%%%%%%%%%%%%%%%%%%%%%%  
Our predictions for $\alpha$ for four $^4$He potentials are given in 
Table~\ref{tab3}. The predictions vary by more than a factor of 2
depending on whether we take $B_2$ or $a$ as the 2-body input.
This large difference arises because the value of $a \Lambda_*$ for
$^4$He atoms is fortuitously close to the value near $a \Lambda_* = 0.83$
at which the sin$^2$ factor in (\ref{3brec-par}) vanishes.
This is illustrated in Fig.~\ref{fig:alpha}, where $\alpha$ in
units of $\hbar a^4/m$ is plotted as a function of $a\Lambda_*$.
%%%%%%%%%%%%%%%%%%%%%%%%%%%%%%%%%%%%%%%%%%%%%%%%%%%%%%%%%%
%%%%%%%%%%%%%%%%%%%%%%%%%%%%%%%%%%%%%%%%%%%%%%%%%%%%%%%%%%%%%% 
The vertical dashed and dot-dashed lines indicate the
values of $a_B\Lambda_*$ and $a\Lambda_*$ for the TTY
potential, respectively. If we use $a$ as input instead of
$B_2$, the sin$^2$ factor is larger by a factor of two.
Because $\alpha$ is positive definite and because it is so 
sensitive to the precise value of $a\Lambda_*$,
we take the universality prediction
to be the geometric mean of the predictions obtained by using
either $B_2$ or $a$ as the input.
We take the theoretical uncertainty to be a multiplicative factor 
equal to the ratio of the two predictions.
The resulting leading order universality prediction for the TTY
potential is
\beq                                        
\mbox{LO}:\qquad \alpha = \left( 0.7^{+1.2}_{-0.4} \right )
        \times 10^{-27} \; {\rm cm}^6/{\rm s} 
        \qquad\qquad\mbox{(TTY)}\,.
\label{eq:alpha_up}
\eeq

There have been several previous calculations of the
3-body recombination rate at threshold for $^4$He.
Fedichev, Reynolds, and Shlyapnikov \cite{FRS96}
calculated the rate by making a semi-analytic approximation
to the Faddeev equations in the hyperspherical representation.
In the limit of large scattering length, they obtained a 
result that depends on $a$ only: $\alpha = 3.9\, \hbar a^4/m$.
They did not observe the oscillatory dependence 
of $\alpha$ on $\ln(a)$ predicted by Efimov theory,
so there must have been an error in their analysis.
They found that solving the Faddeev equations numerically
for the TTY potential for $^4$He gave corrections of about 10\%.
Inserting the value of $a$ for the TTY potential into their analytic
result, we obtain a prediction $\alpha = 0.6 \times 10^{-27}$  cm$^6$/s 
that agrees with the universality prediction in (\ref{eq:alpha_up}).
This agreement is probably fortuitous.

Nielsen and Macek \cite{NiM99}
calculated the 3-body recombination rate at threshold
by applying hidden crossing theory to the Faddeev equations
in the hyperspherical representation.
They found that in the limit of large scattering length,
$\alpha$ could take any value between 0 and $68\, \hbar a^4/m$
depending on some WKB phase. This is consistent with the 
effective field theory result in (\ref{3brec-par}).
For a Gaussian potential that gives the same scattering length
and effective range as the LM2M2 potential, they obtained the prediction
$\alpha = 1.1 \times 10^{-27}$  cm$^6$/s. They 
pointed out that the result is extremely sensitive to their WKB phase
because it is close to the value for which $\alpha$ vanishes.   

Esry, Greene, and Burke \cite{EGB99} calculated
the 3-body recombination rate at threshold by solving the 
Schr{\"o}dinger equation in hyperspherical coordinates
numerically for many potentials with one or at most a few two-body bound 
states.\footnote{The values of $K_3=6\alpha$ given in 
the \lq\lq Present'' column of Table I in Ref.~\protect\cite{EGB99} 
must be divided by 6 in order to correct for a factor of 6 
error in Eq.~(1) of that paper.}
In the case $a>0$, they found that the recombination rates could be
well approximated by an empirical formula that reduces in the large
$a$ limit to an expression with an oscillatory dependence on $\ln a$
similar to Eq.~(\ref{3brec-par}).
In the case of $^4$He, their result for the HFD-B3-FCI1 
potential \cite{AJM95} is $\alpha = 0.12 \times 10^{-27}$  cm$^6$/s.
A new calculation in the hyperspherical adiabatic representation that
also uses the HFD-B3-FCI1 potential has recently
been carried out \cite{SEG02}. This calculation includes states with
angular momentum $J>0$, so that nonzero energies can be considered.
At threshold, it agrees with the result of Ref.~\cite{EGB99}.

The large uncertainty in the universality prediction for $\alpha$ arises
because the value of $a\Lambda_*$ for the $^4$He potentials lies 
fortuitously close to the zero of Eq.~(\ref{3brec-par}). 
In order to improve on the universality prediction (\ref{3brec-par})
within the effective field theory approach, it would be necessary to
include effective range corrections.
If there was an accurate calculation of $\alpha$ for a potential
for which $B_3^{(1)}$ (or another low-energy three-body observable)
is known, we could use that result to estimate
the effective range corrections for other potentials. The only
calculation of $\alpha$ we consider accurate enough is for the
HFD-B3-FCI1 potential in Ref.~\cite{SEG02}. Unfortunately, we are not
aware of any calculation of $B_3^{(1)}$ for that potential.

\section{Summary and Outlook}
\label{sec:conc}

The universality approach to the three-body problem with large
scattering length was pioneered by Efimov \cite{Efi71}, who emphasized
the qualitative insights it provides. This approach implies a
correlation between all the low-energy three-body observables for
potentials that have a large scattering length. 
The universality approach is also useful as a quantitative tool.
It implies that up to corrections suppressed by $l/a$, all low-energy
three-body observables are determined by $a$ and a single three-body
parameter. A convenient choice for the three-body parameter is
the parameter $\Lambda_*$ introduced in Ref.~\cite{BHK99}, because
the dependence of some observables on $\Lambda_*$ is known analytically.

In order to determine $\Lambda_*$, one three-body observable is
required as input. A convenient choice is the binding energy of
the shallowest Efimov state. Once $\Lambda_*$ is determined,
all other low-energy three-body observables can be predicted. 
We used calculations of the binding energy $B_3^{(1)}$ of the excited 
state of the trimer to determine $\Lambda_*$ for various $^4$He 
potentials. 
We then used universality to calculate the binding energy $B_3^{(0)}$ 
of the ground  state of the trimer, the atom-dimer phase shifts below the 
dimer breakup threshold, and the three-body recombination constant at 
threshold $\alpha$.
We gave explicit expressions for the three-body recombination constant
$\alpha$ in  Eq.~(\ref{3brec-par})
and for the S-wave atom-dimer phase shifts below the breakup
threshold in Eq.~(\ref{kcot-Efimov}). 
We also gave an explicit parametrization for the
universal function $\Delta(\xi )$ that appears in Efimov's equation
(\ref{B3-Efimov}) for the trimer binding energies.

The leading corrections to the universality predictions
come from effective range corrections.
If these corrections are included, there should be a systematic 
improvement in the accuracy of the predictions for all low-energy
observables with errors decreasing to second order in $l/a$.
The effective range corrections have not yet been calculated for the
case of three identical spinless bosons with large scattering length.
We therefore used accurate calculations of $B_3^{(0)}$ and $a_{12}$ for the
HFD-B potential as input to deduce the approximate effective range 
corrections in these observables for the other potentials. 
The resulting theoretical errors are smaller than those for the 
leading-order
universality prediction by about an order of magnitude as expected. 
Comparing with the calculated values of $B_3^{(0)}$ and $a_{12}$, we see 
that the theoretical errors correctly reflect the accuracy of the 
LO and NLO universality predictions.
An actual calculation of the effective range corrections for the 
three-body observables would eliminate the need for using calculations 
of  three-body observables for one potential as additional inputs. 

The leading-order universality predictions presented in this paper were
obtained using the effective field theory of Ref.~\cite{BHK99}, which is
a particularly convenient implementation of the universality
approach for three-body systems. 
More generally, effective field theory provides a framework 
for the model independent description of low-energy phenomena
by exploiting a separation of scales in the system \cite{EFT}.
Using renormalization, all short-distance effects are systematically
absorbed into a few low-energy parameters such as the scattering 
length $a$ and $\Lambda_*$. As a consequence, the dependence on the 
relevant low-energy parameters is explicit, while irrelevant details 
of how their numerical values arise from complicated
short-distance dynamics are omitted. 
Effective field theory allows for systematically improvable
calculations of low-energy observables with well-defined error estimates. 
This method has many applications ranging from particle physics over 
nuclear physics to condensed matter physics
\cite{particleEFT,nucEFT,condEFT}. 

An important open question is how universality is manifested
in the four-body problem. Low-energy four-body
observables must depend on the two-body parameter $a$ and the three-body
parameter $\Lambda_*$. Are any new four-body parameters required
to calculate low-energy four-body observables up to corrections
suppressed by $l/|a|$? There are theoretical arguments in support of
both answers to this question. There
is a renormalization argument for zero-range $\delta$-function two-body
potentials that indicates that a new four-body parameter is required
to calculate four-body binding energies \cite{AFG95}. On the other
hand, a power counting argument within the effective
field theory framework suggests that a four-body parameter
should not be necessary to calculate four-body observables to leading
order in $l/|a|$ \cite{Lepage}. This raises the exciting possibility of
calculating the binding energy of the $^4$He tetramers close to the
four-atom threshold to about 10\% accuracy using $a$
and $B_3^{(1)}$ (or another low-energy three-body observable)
as the only inputs.
There is some circumstantial evidence in favor of this possibility
from the four-body
problem in nuclear physics. There is a correlation called the 
\lq\lq Tjon line'' between the binding energy $B_3$ of the triton
($^3$H nucleus) and the binding energy $B_4$ of the alpha particle
($^4$He nucleus) \protect\cite{Tjo75}.
Calculations of these binding energies using modern nucleon-nucleon
interaction potentials give results that underestimate both 
binding energies but cluster along a line in the $B_3$-$B_4$ plane.
By adding a three-body potential whose strength is adjusted to get
the correct value for $B_3$, one also gets an accurate result for 
$B_4$ (see Ref.~\protect\cite{NKG00} for some recent calculations
with modern nuclear forces). 

The results for $^4$He presented in the paper apply equally well to 
other bosonic atoms with large scattering length as long as the effects of
deep two-body bound states on low-energy observables are
sufficiently small. By definition, a deep bound state has a
binding energy of order $\hbar^2/ml^2$ or larger. If $a<0$, any
dimer is deeply bound. If $a>0$, any dimer other than the shallow dimer
with $B_2 \approx\hbar^2/ma^2$ is deeply bound. 
One qualitative effect of the deep
two-body states is that the Efimov states become resonances,
because they can decay into an atom and a deeply bound dimer.
Deep two-body bound states also provide additional channels for
three-body recombination. Their effects can be particularly dramatic 
for $a<0$ if there is an Efimov state near the three-atom
threshold, because it gives a resonant enhancement of the three-body
recombination rate into deep two-body bound states \cite{EGB99,BrH01}.
The existence
of deep two-body bound states does not affect the universality prediction 
for low-energy observables in the two-body sector. However in the 
three-body sector, it implies that a third parameter in
addition to $a$ and $\Lambda_*$ is required to predict low-energy
observables up to corrections suppressed by $l/|a|$ \cite{BrH01}. This
parameter takes into account the cumulative effects of all the
deep two-body bound states. 
The modification of Efimov's equation for the binding energies was given
in Ref.~\cite{BHK02}.

The universality approach discussed in this paper is not limited
to identical bosons. It can be applied to any three-particle system 
for which at least two of the three pairs have a large scattering length.
Some examples are given in a recent review article \cite{Nie01}.
The universality predictions will depend on the three pair-wise
scattering lengths, the ratios of the masses,
and the three-body parameter $\Lambda_*$.

An especially promising
application of the universality approach is to cold atoms 
in the vicinity of a Feshbach resonance, where the effective scattering
length can be controlled by an external magnetic field \cite{Festh}. 
This creates the exciting possibility of testing the unique oscillatory 
dependence of low-energy three-body observables on the scattering length 
that is predicted by universality. 
In this paper, we have exploited the fact that the various $^4$He
potentials span a small interval of $a\Lambda_*$. Using a Feshbach 
resonance to control the scattering length, it might be possible to
scan through an entire cycle of the oscillatory behavior. Among the dramatic
effects that one may be able to observe are the divergence in $a_{12}$ 
near $a\Lambda_*=4.3$ and the zero of $\alpha$ near $a\Lambda_*=0.83$.

The behavior of cold atoms near a Feshbach resonance is in general a 
complicated coupled-channel problem involving the various hyperfine 
states of the atoms.  However, from the point of view of universality,
the coupled-channel effects introduce no additional complications.
If one is sufficiently close to the resonance and if the energy 
relative to the threshold for one hyperfine state
is small compared to the hyperfine splittings, 
only that hyperfine state needs to be included explicitly.  
The coupled-channel effects can be taken into account through the values of
the low-energy parameters $a$, which diverges at the resonance, 
and $\Lambda_*$, which varies slowly in the neighborhood of the resonance.

The behavior of a Bose-Einstein condensate of atoms with large 
scattering lengths has been studied experimentally by using Feshbach 
resonances to tune the scattering lengths of alkali atoms 
\cite{Fesex}. In the low-density limit $na^3 \ll 1$, the nontrivial
aspects of universality in the three-body sector are reflected in
a small oscillatory dependence of the energy density of the condensate
on $\ln(\Lambda_* n^{1/3})$ \cite{BHM02}. There is a  possibility
that these three-body effects would allow the existence of stable 
homogeneous condensates with large negative
scattering length \cite{Bul99}.
The intermediate density region where $na^3 \sim 1$ but
$nl^3 \ll 1$ is a more difficult problem. It is an open question
whether a condensate in this region has universal properties that are
determined by constants such as $a$ and $\Lambda_*$ that describe
the low-energy properties in the few-body sectors. If there are, it may 
be possible to use universality to predict in detail the behavior
of a Bose-Einstein condensate of atoms near a Feshbach resonance.

\begin{acknowledgments}
We thank D.\ Blume, B.D.\ Esry, and C.H.\ Greene for useful
correspondence.
This research was supported by DOE Grant No. DE-FG02-91-ER4069 and
NSF Grant No. PHY-0098645.
\end{acknowledgments}

\end{document}